\documentclass[aps,prc,letterpaper,11pt,twoside,tightenlines,nofootinbib,showpacs,preprint]{revtex4}
\usepackage{graphicx}
\usepackage{epsfig,float}
\usepackage[sort&compress]{natbib}
\usepackage{subfigure}
\usepackage{amsmath}
\usepackage{amsfonts}
\usepackage{cancel}
\usepackage{lmodern,dsfont}
\usepackage{amssymb}
\usepackage{hyperref}
\begin{document}
\arraycolsep1.5pt
\newcommand{\Ima}{\textrm{Im}}
\newcommand{\Rea}{\textrm{Re}}
\newcommand{\mev}{\textrm{ MeV}}
\newcommand{\gev}{\textrm{ GeV}}
\newcommand{\dtres}{d^{\hspace{0.1mm} 3}\hspace{-0.5mm}}
\newcommand{\rts}{ \sqrt s}
\newcommand{\non}{\nonumber \\[2mm]}
\newcommand{\eps}{\epsilon}
\newcommand{\half}{\frac{1}{2}}
\newcommand{\thalf}{\textstyle \frac{1}{2}}
\newcommand{\Nmass}{M_{N}} 
\newcommand{\delmass}{M_{\Delta}} 
\newcommand{\pimass}{\mu}  
\newcommand{\rhomass}{m_\rho} 
\newcommand{\piNN}{f}      
\newcommand{\rhocoup}{g_\rho} 
\newcommand{\fpi}{f_\pi} 
\newcommand{\f}{f} 
\newcommand{\nucfld}{\psi_N} 
\newcommand{\delfld}{\psi_\Delta} 
\newcommand{\fpiNN}{f_{\pi N N}} 
\newcommand{\fpiND}{f_{\pi N \Delta}} 
\newcommand{\GMquark}{G^M_{(q)}} 
\newcommand{\vecpi}{\vec \pi}
\newcommand{\vectau}{\vec \tau}
\newcommand{\vecrho}{\vec \rho}
\newcommand{\delmu}{\partial_\mu}
\newcommand{\delMu}{\partial^\mu}
\newcommand{\nn}{\nonumber}
\newcommand{\bi}{\bibitem}
\newcommand{\vs}{\vspace{-0.20cm}}
\newcommand{\be}{\begin{equation}}
\newcommand{\ee}{\end{equation}}
\newcommand{\ba}{\begin{eqnarray}}
\newcommand{\ea}{\end{eqnarray}}
\newcommand{\ropi}{$\rho \rightarrow \pi^{0} \pi^{0}
\gamma$ }
\newcommand{\roeta}{$\rho \rightarrow \pi^{0} \eta
\gamma$ }
\newcommand{\omepi}{$\omega \rightarrow \pi^{0} \pi^{0}
\gamma$ }
\newcommand{\omeeta}{$\omega \rightarrow \pi^{0} \eta
\gamma$ }
\newcommand{\ul}{\underline}
\newcommand{\del}{\partial}
\newcommand{\rth}{\frac{1}{\sqrt{3}}}
\newcommand{\rsix}{\frac{1}{\sqrt{6}}}
\newcommand{\sq}{\sqrt}
\newcommand{\fr}{\frac}
\newcommand{\pr}{^\prime}
\newcommand{\ov}{\overline}
\newcommand{\Gm}{\Gamma}
\newcommand{\rw}{\rightarrow}
\newcommand{\rgl}{\rangle}
\newcommand{\De}{\Delta}
\newcommand{\Dp}{\Delta^+}
\newcommand{\Dm}{\Delta^-}
\newcommand{\Dz}{\Delta^0}
\newcommand{\Dpp}{\Delta^{++}}
\newcommand{\Sg}{\Sigma^*}
\newcommand{\Sp}{\Sigma^{*+}}
\newcommand{\Sm}{\Sigma^{*-}}
\newcommand{\Sz}{\Sigma^{*0}}
\newcommand{\X}{\Xi^*}
\newcommand{\Xm}{\Xi^{*-}}
\newcommand{\Xz}{\Xi^{*0}}
\newcommand{\Om}{\Omega}
\newcommand{\Omm}{\Omega^-}
\newcommand{\kp}{K^+}
\newcommand{\kz}{K^0}
\newcommand{\pip}{\pi^+}
\newcommand{\pim}{\pi^-}
\newcommand{\piz}{\pi^0}
\newcommand{\et}{\eta}
\newcommand{\kb}{\ov K}
\newcommand{\km}{K^-}
\newcommand{\kbz}{\ov K^0}
\newcommand{\ksb}{\ov {K^*}}

\def\tstrut{\vrule height2.5ex depth0pt width0pt} 
\def\jtstrut{\vrule height5ex depth0pt width0pt} 

\title{Interaction of vector mesons with baryons and nuclei}

\author{E. Oset$^{1}$, A. Ramos$^2$, E. J. Garzon$^1$, R. Molina$^3$, L. Tolos$^4,5$, C. W. Xiao$^1$, J. J. Wu$^6$ and B. S. Zou$^{7,8}$}
\affiliation{
$^{1}$Departamento de F\'{\i}sica Te\'orica and IFIC, Centro Mixto Universidad de 
Valencia-CSIC,
Institutos de Investigaci\'on de Paterna, Aptdo. 22085, 46071 Valencia,
Spain\\
$^2$ Departament d'Estructura i Constituents de la Materia, Universitat de
Barcelona \\
$^3$Research Center for Nuclear Physics (RCNP),
Mihogaoka 10-1, Ibaraki 567-0047, Japan\\
$^4$ Instituto de Ciencias del Espacio (IEEC/CSIC) Campus Universitat Aut\`onoma de Barcelona, Facultat de Ci\`encies, Torre C5, E-08193 Bellaterra (Barcelona),  Spain\\
$^5$ Frankfurt Institute for Advanced Studies (FIAS). Johann Wolfgang Goethe University. Ruth-Moufang-Str. 1. 60438 Frankfurt am Main. Germany\\
$^6$Physics Division, Argonne National Laboratory, Argonne, Illinois 60439, USA\\
$^7$Institute of High Energy Physics, CAS, P.O.Box 918(4), Beijing 100049, China\\
$^8$Theoretical Physics Center for Science Facilities, CAS, Beijing 100049, China
 }

\date{\today}

\begin{abstract}

After some short introductory remarks on particular issues on the vector mesons in nuclei, in this paper we present a short review of recent developments concerning the interaction of vector mesons with baryons and with nuclei from a modern perspective using the local 
hidden gauge formalism for the interaction of vector mesons. We present results for the vector baryon interaction and in particular for the resonances which appear as composite states, dynamically generated from the interaction of vector mesons with baryons, taking also the mixing of these states with pseudoscalars and baryons into account. We then venture into the charm sector, reporting on hidden charm baryon states around 4400 MeV, generated from the interaction of vector mesons and baryons with charm, which have a strong repercussion on the properties of the 
$J/\Psi N$ interaction. 
 We also address the interaction of $K^*$ with nuclei and make suggestions to measure the predicted huge width in the medium by means of the transparency ratio. The formalism is extended to study the phenomenon of $J/\psi$ suppression in nuclei via $J/\psi$ photoproduction reactions.
\end{abstract}

\maketitle

\section{Introduction}
\label{Intro}

The topic of vector meson interactions with nuclei has attracted, and continues to attract, much attention. After thorough experimental and theoretical studies of pion nuclear interaction and other pseudoscalar nucleus interactions, the turn came for exploring the properties of the vector mesons in nuclei. In this limited study we do not pretend to make an exhaustive review of the field, which has been done anyway in other papers \cite{rapp} and more recently in \cite{Hayano:2008vn,Leupold:2009kz}. We shall make emphasis on the new perspective that the use of the local hidden gauge theory \cite{hidden1,hidden2,hidden4} has brought into this topic.

   From a historical perspective, one certainly must admit, that in spite of much theoretical evidence against it from detailed calculations \cite{Rapp:1997fs,Peters:1997va,Rapp:1999ej,Urban:1999im,Cabrera:2000dx,Cabrera:2002hc}, the guess in \cite{Brown:1991kk} that the vector meson masses would be drastically reduced in a nuclear medium stimulated much experimental work. Experimental searches have finally concluded that this was not the case \cite{na60,wood,Hayano:2008vn,Leupold:2009kz}. However,  a few surprises have been found on the way. 
    
    As a matter of example let us recall here the history concerning the $\omega$ mass in the medium from experiments in the ELSA (Bonn) Laboratory. The analysis of the experimental results on the photoproduction of $\omega$, observed from the decay channel $\pi^0 \gamma$, led the authors of \cite{Trnka:2005ey}
to claim the first observation of in-medium modifications of the $\omega$ meson mass, by an approximate amount of 100 MeV for normal nuclear matter density. Yet, the observation in \cite{Kaskulov:2006zc} that the results of \cite{Trnka:2005ey} were tied to a particular choice of background led to a thorough search for background processes in \cite{Nanova:2010sy} that finished with the withdrawal of the $\omega$ mass shift claims. In between, suggestions that some signal reported in \cite{tesistrnka,Metag:2007zz} could be indicative of an $\omega$ meson bound state in nuclei were aborted very early, realizing that a double hump structure in the experiment was due to a different scaling of the uncorrelated $\pi^0 \gamma$ production events and the $\omega$ production process with subsequent $\pi^0 \gamma$ decay \cite{Kaskulov:2006fi}. Before the final conclusions of \cite{Nanova:2010sy}, it was also hoped that the use of the successful mixed events method to separate the background from the signal should be sufficient to isolate the $\omega$ signal \cite{Metag:2007hq}. However, a simulation of the reaction \cite{mixed} showed that
the mixed event method produced a background essentially independent of the real background in the region of relevance to omega production, basically determined from events occurring at much lower invariant masses.

 The persistence of both theoretical and experimental teams in the clarification of the problem gave undoubtedly some fruits and this is now the problem most thoroughly studied in this field, that has led to clarifications in other related problems. Yet, there is some interesting physical information that  survived the close scrutiny of the former works and this is the large width of the $\omega$ in the medium found in \cite{Kotulla:2008aa} and also studied in \cite{Kaskulov:2006zc}. The width of the $\omega$ in the medium extrapolated to normal nuclear matter density was of the order of 100 MeV in \cite{Kaskulov:2006zc} and 130-150 MeV in \cite{Kotulla:2008aa}. The theoretical understanding of this large width, related to decay channels of the $\omega$ in the nuclear medium, is a challenge that will require combined efforts of hadron dynamics and many body theory.

In this paper we shall review recent developments on the interaction of vector mesons with baryons and nuclei, using effective field theory with
a combination of effective Lagrangians to account for hadron interactions, and  implementing exactly unitarity in coupled channels. This approach is a very efficient tool to 
 face many problems in Hadron Physics. Using this  coupled channel unitary approach with the input from chiral Lagrangians, usually referred to as chiral unitary approach, the interaction of the octet of
pseudoscalar mesons with the octet of stable baryons has been studied and 
leads to $J^P=1/2^-$
resonances which fit quite well the spectrum of the known low lying resonances
with these quantum numbers 
\cite{Kaiser:1995cy,angels,ollerulf,carmenjuan,hyodo,ikeda}. 
Among the new resonances predicted, the most notable is the
$\Lambda(1405)$, where all the chiral approaches find two poles close by 
\cite{Jido:2003cb,Borasoy:2005ie,Oller:2005ig,Oller:2006jw,Borasoy:2006sr,Hyodo:2008xr,Roca:2008kr}, rather than one, for which 
experimental support is presented in \cite{magas,sekihara}.  Another step forward in this direction has been the interpretation
of low lying $J^P=1/2^+$ states as molecular systems of two pseudoscalar mesons and one baryon
\cite{alberto,alberto2,kanchan,Jido:2008zz,KanadaEn'yo:2008wm}. 

More recently, vectors instead of
pseudoscalars have also been considered. In the baryon sector the
interaction of the $\rho \Delta$ interaction was addressed in
\cite{vijande}, where three degenerate $N^*$ states and three degenerate
$\Delta$ states around 1900 MeV, with $J^P=1/2^-, 3/2^-, 5/2^-$, were found. The extrapolation to SU(3) with the interaction of the vectors of the nonet with
the baryons of the decuplet was studied in \cite{sourav}. The starting point
of these works is the hidden gauge formalism
\cite{hidden1,hidden2,hidden4}, which deals with the interaction of vector mesons and
pseudoscalars, respecting chiral dynamics, providing the interaction of
pseudoscalars among themselves, with vector mesons, and vector mesons among
themselves. It also offers a perspective on the chiral Lagrangians as limiting
cases at low energies of vector exchange diagrams occurring in the theory.

  The results of the interaction of the nonet of vector mesons  with the
   octet of baryons were reported in \cite{angelsvec}. The scattering amplitudes, obtained under the
approximation of neglecting the three momentum of the particles versus their
mass, led to poles in the
complex plane which can be associated to some well known resonances. In \cite{angelsvec} one obtains degenerate states of $J^P=1/2^-,3/2^-$, a degeneracy that
seems to be followed qualitatively by the experimental spectrum, although in
some cases the spin partners have not been identified. We will also report on improvements in this theory which
consider the mixing of the vector-baryon states with pseudoscalar-baryon ones.

In fact, there is in principle no reason to expect that the interaction of
pseudoscalar mesons with baryons and the interaction of vector mesons
with baryons should be decoupled for states which share strangeness,
isospin, and $J^P$ (spin-parity) quantum numbers. The consequences of
coupling these interactions, that have been treated independently in
the previous works of \cite{sourav} and \cite{angelsvec},
were first explored in the three flavour sector in
Ref.~\cite{Gamermann:2011mq}. In that work, an SU(6)
framework ~\cite{GarciaRecio:2005hy,GarciaRecio:2006wb,Toki:2007ab}
was used which combines spin and
flavor symmetries within an enlarged Weinberg-Tomozawa meson--baryon
Lagrangian in order to accommodate vector mesons and decuplet baryons. This
guarantees that chiral symmetry is recovered when interactions
involving pseudoscalar Nambu-Goldstone bosons are being
examined\footnote{A similar study for the case of
meson-meson light resonances was carried out in
Ref.~\cite{GarciaRecio:2010ki}.}.
Chiral symmetry constraints the pseudoscalar octet--baryon decuplet
interactions. However, the interaction of vector mesons with baryons is not
constrained by chiral symmetry, and thus the model presented
in \cite{Gamermann:2011mq}  differs
from those of Refs.~\cite{sourav,angelsvec}, based on the hidden gauge
formalism. However, in the presence of heavy quarks the analogous
scheme to that of Ref.~\cite{Gamermann:2011mq} automatically embodies
heavy quark spin symmetry, another well established approximate
symmetry of QCD.  Indeed, the model of Ref.~\cite{Gamermann:2011mq}
has been successfully extended to the  charm sector
in \cite{GarciaRecio:2008dp,Gamermann:2010zz,Romanets:2012hm}. 

The vast number of resonances with charm or hidden charm found in the recent years, some of which having a clear molecular structure, and the possibility of studying them copiously at LHC or upgraded B-factories, has injected a
renewed interest in this field. We will report on composite states of hidden charm emerging from the interaction of vector mesons and baryons with charm.  

 Finally, we devote some attention to new developments on the properties of  vector mesons in a nuclear medium, focusing specifically on the interactions of the $K^*$ and $J/\psi$ mesons with nuclei.

\section{Formalism for $VV$ interaction}
\label{sec:formalism}

The formalism of the hidden gauge interaction for vector mesons is provided in 
\cite{hidden1,hidden2} (see also \cite{hidekoroca} for a practical set of Feynman rules). 
The Lagrangian involving the interaction of 
vector mesons amongst themselves is given by
\begin{equation}
{\cal L}_{III}=-\frac{1}{4}\langle V_{\mu \nu}V^{\mu\nu}\rangle \ ,
\label{lVV}
\end{equation}
where the symbol $\langle \rangle$ stands for the trace in the SU(3) space 
and $V_{\mu\nu}$ is given by 
\begin{equation}
V_{\mu\nu}=\partial_{\mu} V_\nu -\partial_\nu V_\mu -ig[V_\mu,V_\nu]\ ,
\label{Vmunu}
\end{equation}
with  $g$ given by $g=\frac{M_V}{2f}$
where $f=93$~MeV is the pion decay constant. The magnitude $V_\mu$ is the SU(3) 
matrix of the vectors of the nonet of the $\rho$
\begin{equation}
V_\mu=\left(
\begin{array}{ccc}
\frac{\rho^0}{\sqrt{2}}+\frac{\omega}{\sqrt{2}}&\rho^+& K^{*+}\\
\rho^-& -\frac{\rho^0}{\sqrt{2}}+\frac{\omega}{\sqrt{2}}&K^{*0}\\
K^{*-}& \bar{K}^{*0}&\phi\\
\end{array}
\right)_\mu \ .
\label{Vmu}
\end{equation}

The interaction of ${\cal L}_{III}$ gives rise to a contact term coming from 
$[V_\mu,V_\nu][V_\mu,V_\nu]$
\begin{equation}
{\cal L}^{(c)}_{III}=\frac{g^2}{2}\langle V_\mu V_\nu V^\mu V^\nu-V_\nu V_\mu
V^\mu V^\nu\rangle\ ,
\label{lcont}
\end{equation}
as well as to a three 
vector vertex from 
\begin{equation}
{\cal L}^{(3V)}_{III}=ig\langle (\partial_\mu V_\nu -\partial_\nu V_\mu) V^\mu V^\nu\rangle
=ig\langle (V^\mu\partial_\nu V_\mu -\partial_\nu V_\mu
V^\mu) V^\nu\rangle
\label{l3Vsimp}\ .
\end{equation}

It is worth stressing the analogy with the coupling of vectors to
 pseudoscalars given in the same theory by  
\be
{\cal L}_{VPP}= -ig ~\langle [P,\partial_{\mu}P]V^{\mu}\rangle,
\label{lagrVpp}
\ee
where $P$ is the SU(3) matrix of the pseudoscalar fields. 

The Lagrangian for the coupling of vector mesons to
the baryon octet is given by
\cite{Klingl:1997kf,Palomar:2002hk} 
\be
{\cal L}_{BBV} =
\frac{g}{2}\left(\langle\bar{B}\gamma_{\mu}[V^{\mu},B]\rangle+\langle\bar{B}\gamma_{\mu}B\rangle \langle V^{\mu}\rangle \right),
\label{lagr82}
\ee
where $B$ is now the SU(3) matrix of the baryon octet \cite{Eck95,Be95}. Similarly,
one has also a Lagrangian for the coupling of the vector mesons to the baryons
of the decuplet, which can be found in \cite{manohar}.

\begin{figure}[tb]
\epsfig{file=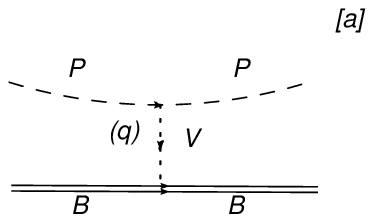, width=7cm} \epsfig{file=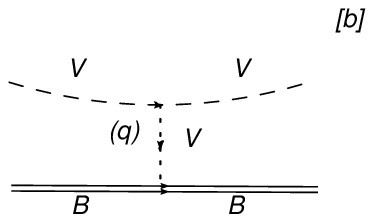, width=7cm}
\caption{Diagrams obtained in the effective chiral Lagrangians for interaction
of pseudoscalar [a] or vector [b] mesons with the octet or decuplet of baryons.}%
\label{f1}%
\end{figure}

Starting from these Lagrangians one can draw the Feynman diagrams that lead to the $PB
\to PB$ and $VB \to VB$ interaction, by exchanging a vector meson between the
pseudoscalar or the vector meson and the baryon, as depicted in Fig.\ref{f1} .

 It was shown in \cite{angelsvec} that in the limit of small three momenta of the vector mesons the vertices of Eq. (\ref{l3Vsimp}) and Eq. (\ref{lagrVpp}) give rise to the same expression.  This makes the work technically easy, allowing the use of many previous results.

   A caveat must be made in the case of vector mesons due to the mixing of $\omega_8$ and the singlet of SU(3), $\omega_1$, to give the
   physical states of the $\omega$ and the $\phi$.
    In this case, all one must do is to take the
   matrix elements known for the $PB$ interaction and, wherever $P$ is the
   $\eta_8$, multiply the amplitude by the factor $1/\sqrt 3$ to get the
   corresponding $\omega $ contribution and by $-\sqrt {2/3}$ to get the
   corresponding $\phi$ contribution.  Upon the approximation consistent with
   neglecting the three momentum versus the mass of the particles (in this
   case the baryon), we can just take the $\gamma^0$ component of 
   Eq. (\ref{lagr82})  and
   then the transition potential corresponding to the diagram of Fig. 1(b) is
   given by
   \begin{equation}
V_{i j}= - C_{i j} \, \frac{1}{4 f^2} \, (k^0 + k'^0)~ \vec{\epsilon}\vec{\epsilon
} ',
\label{kernel}
\end{equation}
 where $k^0, k'^0$ are the energies of the incoming and outgoing vector mesons. 
   The same occurs in the case of the decuplet.    
    The $C_{ij}$ coefficients of Eq. (\ref{kernel}) can be obtained directly from 
    \cite{angels,bennhold,inoue}
    with the simple rules given above for the $\omega$ and the $\phi$, and
    substituting $\pi$ by $\rho$ and $K$ by $K^*$ in the matrix elements.

The scattering matrix is constructed by solving the
    coupled channels Bethe Salpeter equation in the on shell factorization approach of 
    \cite{angels,ollerulf}
   \begin{equation}
T = [1 - V \, G]^{-1}\, V,
\label{eq:Bethe}
\end{equation} 
with $G$ being the loop function of a vector meson and a baryon, which we calculate in
dimensional regularization using the formula of \cite{ollerulf} and similar
values for the subtraction constants. In the present case the $\rho$ and the $K^*$ have a significant width and the $G$ functions involving these mesons must be convoluted with their corresponding
spectral functions.

 The iteration of diagrams implicit in the Bethe Salpeter equation in the case
 of the vector mesons propagates the $\vec{\epsilon}\vec{\epsilon }'$ term 
 of the interaction. Hence,
the factor $\vec{\epsilon}\vec{\epsilon }'$ appearing in the potential $V$
factorizes also in the $T$ matrix for the external vector mesons. As a consequence of this, the interaction is spin independent and one finds degenerate states having $J^P=1/2^-$ and $J^P=3/2^-$.

\section{Incorporating the pseudoscalar meson-baryon channels}

\begin{center}
\begin{figure}[h!]
\begin{center}
\includegraphics[scale=0.5]{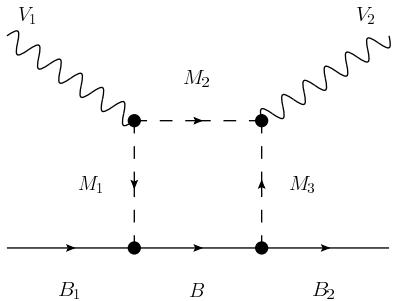} 
\end{center}
\caption{ Diagram for the $VB \rightarrow VB$ interaction incorporating the intermediate pseudoscalar-baryon states.}
\label{box}
\end{figure}
\end{center}

\begin{figure}[ht!]
\begin{center}
\includegraphics[scale=0.5]{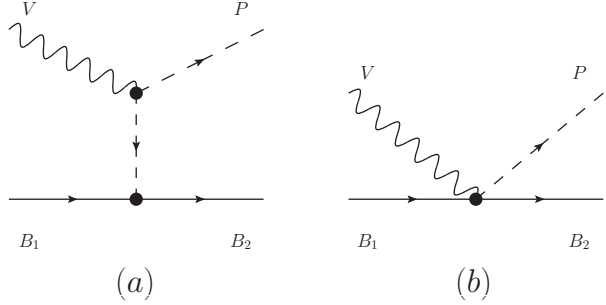}
\end{center}
\caption{Diagrams of the $VB\rightarrow PB$. (a) meson pole term, (b) Kroll-Ruderman contact term.}
\label{fig:vbpb}
\end{figure}

Improvements to the work of \cite{angelsvec} have been done by incorporating intermediate states of a pseudoscalar meson and a baryon in \cite{garzon}. This is practically implemented by including the diagrams of Fig.~(\ref{box}). However, arguments of gauge invariance \cite{Rapp:1997fs,Peters:1997va,Rapp:1999ej,Urban:1999im,Cabrera:2000dx,Cabrera:2002hc,kanchan1,kanchan2} demand that the meson pole term be accompanied by the corresponding Kroll-Ruderman contact term, see Fig. \ref{fig:vbpb}.

In the intermediate $B$ states of Fig. \ref{box} we include baryons of the octet and the decuplet. The results of the calculations are a small shift and a broadening of the resonances compared with what is obtained with the basis of vector-baryon alone. 
  
\begin{figure}[ht!]
\begin{center}
\includegraphics[scale=1.2]{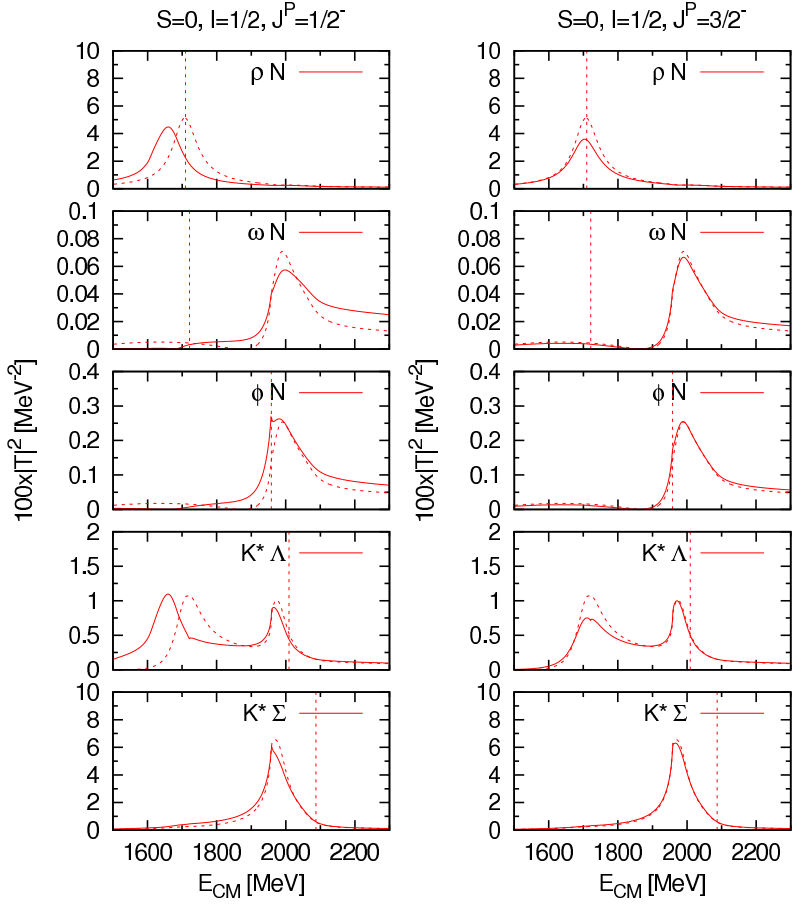} 
\end{center}
\caption{$|T|^2$ for the S=0, I=1/2 states. Dashed lines correspond to tree level only and solid lines are calculated including the box diagram potential. Vertical dashed lines indicate the channel threshold.}
\label{res1}
\end{figure}

In Fig.~\ref{res1} we observe two peaks for the states having quantum numbers $S=0$ and $I=1/2$, one around 1700 MeV, in channels $\rho N$ and $K^* \Lambda$, and another peak near 1980 MeV, which appears in all the channels except for $\rho N$. 
It can be seen that the mixing of the $PB$ channels (solid lines) affects differently the two spin sectors, $J^P=1/2^-$ and $3/2^-$, as a consequence of the extra mechanisms contributing mainly to the $J^P=1/2^-$. 
Indeed, the $PB-VB$ mixing mechanism is more important in the $J^P=1/2^-$ sector because the Kroll-Ruderman term only allows the $1/2^-$ pseudoscalar baryon intermediate states in the box diagram.
 The most important feature is a breaking of the degeneracy for the peak around 1700 MeV. This is most welcome since allows us to associate the $1/2^-$ peak found at 1650 MeV with the $N^*(1650)(1/2^-)$ while the peak for $3/2^-$ at 1700 MeV can be naturally associated to the $N^*(1700)(3/2^-)$. However, let us recall that in the baryon lines of Fig. \ref{box} we only include ground states ($N$ and $\Delta$). A resonance like the $N^*(1520)(3/2^-)$ also appears dynamically generated in the scheme extending the space to $\pi N$(d-wave) and $\pi \Delta$(d-wave) and will be reported elsewhere \cite{javinew}.

 The resonances obtained are summarized in Tables \ref{tab:pdg12} and \ref{tab:pdg32}..
There are states which one can easily
associate to known resonances, and there remain ambiguities in other cases. The nearly degeneracy in spin that the theory predicts is
clearly visible in the experimental data, as one sees a few states with about 50 MeV or less mass difference
between them.  In some cases, the theory predicts quantum numbers for 
resonances which have no assigned spin and parity. It would be interesting to
pursue the experimental determination of these quantum numbers to test the theoretical predictions. 
In addition, the predictions made here for resonances not yet observed should be a stimulus for further search of such states. In this
sense it is worth noting the experimental program at Jefferson Lab 
\cite{Price:2004xm} to investigate the $\Xi$ resonances. With admitted uncertainties of about 50 MeV in masses, we are
confident that the predictions shown here stand on solid grounds and look forward to
progress in the area of baryon spectroscopy and on the understanding of the
nature of the baryonic resonances. 

Before finishing this section we note that the vector meson-baryon and pseudoscalar-baryon coupled problem has also been studied by taking the s-, t- and u-channel diagrams together with a contact term originating from the hidden local symmetry Lagrangian to obtain the $VB$ interaction \cite{xx1}. The $PB$ interaction was calculated using the Kroll-Ruderman contact term, which contrary to \cite{garzon}, refrained the authors from investigating the possible effects of coupling $PB$ to $VB$ channels with spin 3/2. In the spin 1/2 case, some new resonances were found to get generated by the coupled $PB$ and $VB$ dynamics \cite{xx2,kanchan1,kanchan2}. The coupling constants of the low-lying resonances to the $VB$ channels were also obtained in Ref.~\cite{xx2,kanchan1}, which can be useful in studying the photoproduction of these states.

\begin{table}[ht]
\begin{center}
\begin{tabular}{c|c|cc|ccccc}
\hline\hline
$S,\,I$	&\multicolumn{3}{c|}{Theory} & \multicolumn{5}{c}{PDG data}\\
\hline
    	& pole position	& \multicolumn{2}{c|}{real axis} &  &  & &  &  \\
    	& $M_R+i\Gamma /2$	& mass & width &name & $J^P$ & status & mass & width \\
\hline
$0,1/2$ & $1690+i24^{*}			$	& 1658  & 98  
		& $N(1650)$ & $1/2^-$ & $\star\star\star\star$ 	& 1645-1670	& 145-185\\
  		
       	& $1979+i67			$	& 1973 & 85 
   		& $N(2090)$ & $1/2^-$ & $\star$ 	 & $\approx 2090$ & 100-400 \\
\hline
$-1,0$	& $1776+i39				$	& 1747 & 94  
  		& $\Lambda(1800)$ & $1/2^-$ & $\star\star\star$ 		& 1720-1850 & 200-400 \\
  		
        & $1906+i34^{*}			$ 	& 1890 & 93  
        & $\Lambda(2000)$ & $?^?$ & $\star$ 					& $\approx 2000$ & 73-240\\

        & $2163+i37				$ 	& 2149 & 61  &  &  &  & & \\
\hline
$-1,1$  & $ -			$	& 1829& 84  
		& $\Sigma(1750)$ & $1/2^-$ & $\star\star\star$ & 1730-1800 & 60-160 \\
   		& $ -			 $ 	& 2116 & 200-240  
   		& $\Sigma(2000)$ & $1/2^-$ & $\star$ 			& $\approx 2000$ & 100-450 \\
\hline
$-2,1/2$& $2047+i19^{*}	$	& 2039 & 70  
		& $\Xi(1950)$ & $?^?$ & $\star\star\star$ 	& $1950\pm15$ & $60\pm 20$ \\

        & $ -			$	& 2084 & 53   
        & $\Xi(2120)$ & $?^?$ & $\star$ 			& $\approx 2120$ & 25  \\
\hline\hline
\end{tabular}
\caption{The properties of the nine dynamically generated resonances and their possible PDG
counterparts for $J^P=1/2^-$. The numbers with asterisk in the imaginary part of the pole position are obtained without convoluting with the vector mass distribution of the $\rho$ and $K^*$.}
\label{tab:pdg12}
\end{center}
\end{table}

\begin{table}[ht]
\begin{center}
\begin{tabular}{c|c|cc|ccccc}
\hline\hline
$S,\,I$	&\multicolumn{3}{c|}{Theory} & \multicolumn{5}{c}{PDG data}\\
\hline
    	& pole position	& \multicolumn{2}{c|}{real axis} &  &  & &  &  \\
    	& $M_R+i\Gamma /2$	& mass & width &name & $J^P$ & status & mass & width \\
\hline
$0,1/2$ & $1703+i4^{*}			$	& 1705  & 103  
  		& $N(1700)$ & $3/2^-$ & $\star\star\star$ 		& 1650-1750 & 50-150\\
  		
       	& $1979+i56			$	& 1975 & 72 
       	& $N(2080)$ & $3/2^-$ & $\star\star$ & $\approx 2080$ & 180-450 \\	
\hline
$-1,0$	& $1786+i11				$	& 1785 & 19  
		& $\Lambda(1690)$ & $3/2^-$ & $\star\star\star \star$ 	& 1685-1695 & 50-70 \\
  		
        & $1916+i13^{*}			$ 	& 1914 & 59  
        & $\Lambda(2000)$ & $?^?$ & $\star$ 					& $\approx 2000$ & 73-240\\

        & $2161+i17				$ 	& 2158 & 29  &  &  &  & & \\
\hline
$-1,1$  & $ -			$	& 1839& 58  
  		& $\Sigma(1940)$ & $3/2^-$ & $\star\star\star$ & 1900-1950 & 150-300\\
   		& $ -			 $ 	& 2081 & 270  &  &  &  & & \\  
\hline
$-2,1/2$& $2044+i12^{*}	$	& 2040 & 53  
		& $\Xi(1950)$ & $?^?$ & $\star\star\star$ 	& $1950\pm15$ & $60\pm 20$ \\

        & $2082+i5^{*} 	$	& 2082 & 32   
        & $\Xi(2120)$ & $?^?$ & $\star$ 			& $\approx 2120$ & 25  \\
\hline\hline
\end{tabular}
\caption{The properties of the nine dynamically generated resonances and their possible PDG
counterparts for $J^P=3/2^-$. The numbers with asterisk in the imaginary part of the pole position are obtained without convoluting with the vector mass distribution of the $\rho$ and $K^*$.}
\label{tab:pdg32}
\end{center}
\end{table}

\section{Hidden charm baryons from vector-baryon interaction}
Following the idea of \cite{angelsvec} it was found in \cite{wu} that several baryon states emerged as hidden charm composite states of mesons and baryons with charm. In particular, in the context of vector-baryon interaction, a hidden charm baryon that couples to $J/\psi N$ and other related channels was found, as shown in Tables \ref{jpsicoupling} and \ref{jpsiwidth}.  This will play a role later on when we discuss the $J/\psi$ suppression in nuclei. The calculations in the charm sector require an extension to SU(4) of the hidden gauge Lagrangians, but the symmetry is explicitly broken by using the physical masses of the hadrons involved in the processes. Therefore, when the exchange of a heavy vector meson is implied, the appropriate reduction in the Feynman diagram is taken into account. 

In section \ref{supp} we report on the study of $J/\psi$ propagation in nuclei. 

                                                                                   \begin{table}[ht]
      \renewcommand{\arraystretch}{1.1}
     \setlength{\tabcolsep}{0.4cm}
\begin{center}
\begin{tabular}{cccccc}\hline
$(I, S)$&  $z_R$                   & \multicolumn{4}{c}{$g_a$}\\
\hline
$(1/2, 0) $   &             & $\bar{D}^{*} \Sigma_{c}$ & $\bar{D}^{*} \Lambda^{+}_{c}$&  $J/\psi N$ \\
          & $4415-9.5i$   & $2.83-0.19i     $          &$-0.07+0.05i   $            &  $-0.85+0.02i$ \\
          &             &$ 2.83  $                 &$0.08  $                  &  $0.85$ \\
\hline
$(0, -1) $ &                & $\bar{D}^{*}_{s} \Lambda^{+}_{c}$   & $\bar{D}^{*} \Xi_{c}$ & $\bar{D}^{*} \Xi'_{c}$ & $J/\psi \Lambda$\\
       &   $4368-2.8i  $  & $1.27-0.04i     $                     &$ 3.16-0.02i $          & $-0.10+0.13i  $          & $0.47+0.04i   $     \\
       &                & $1.27 $                             & $3.16 $               & $0.16 $                & $0.47  $          \\
       &   $4547-6.4i $   & $0.01+0.004i$                         & $0.05-0.02i$            & $2.61-0.13i $            & $-0.61-0.06i   $                \\
       &                & $0.01   $                           & $0.05 $               & $2.61$                 & $0.61 $                  \\
\hline\end{tabular} \caption{Pole position ($z_R$) and coupling
constants ($g_a$) to various channels for the states from
$VB\rightarrow VB$ including the $J/\psi N$ and $J/\psi\Lambda$
channels. }
\label{jpsicoupling}
\end{center}
       \renewcommand{\arraystretch}{1.1}
     \setlength{\tabcolsep}{0.4cm}
\begin{center}
\begin{tabular}{ccccc}\hline
$(I, S)$      &  $z_R$      & \multicolumn{2}{c}{Real axis} & $\Gamma_i$ \\
          &    & $M$ & $\Gamma$                  & \\
\hline
$(1/2, 0)$    &            &      &             &  $J/\psi N$\\
          & $4415-9.5i$  & $4412$ & $47.3$        &  $19.2$       \\
\hline
$(0, -1)$     &            &      &             &  $J/\psi\Lambda$\\
          & $4368-2.8i$  & $4368 $& $28.0 $       &  $5.4$        \\
          & $4547-6.4i $ & $4544 $& $36.6 $       &  $13.8$        \\
\hline\end{tabular} \caption{Pole position ($z_R$), mass ($M$),
total width ($\Gamma$, including the contribution from the light
meson and baryon channel) and the decay widths for the $J/\psi N$
and $J/\psi\Lambda$ channels ($\Gamma_i$). The units are in MeV.}
\label{jpsiwidth}
\end{center}
\end{table}  

\section{The properties of $K^*$ in nuclei}
Much work about the vector mesons $\rho,\phi,\omega$ in nuclei has been done looking for dileptons
\cite{na60,wood,hades,buss,naruki}. Maybe this technical detail is what has prevented any attention being directed to the renormalization of the  $K^*$ in nuclei. However, recently this problem has been addressed in \cite{lauraraquel} with very interesting results. 


The $K^{*-}$ width in vacuum is determined in \cite{lauraraquel} by evaluating the imaginary part of the free $\bar K^*$ self-energy at rest,  ${\rm Im} \Pi^0_{\bar K^*}$, due to the decay of the $\bar{K}^*$ meson into $\bar{K}\pi$ pairs, using the model parameters of the Lagrangians described in Sect.~\ref{sec:formalism}. The obtained width, $\Gamma_{K^{*-}}=-\mathrm{Im}\Pi_{\bar{K}^*}^{0}/m_{\bar K^*}=42$~MeV,  is quite close to the experimental value of $50.8\pm 0.9$ MeV.

\begin{figure}[t]
\begin{center}
\includegraphics[width=0.45\textwidth,height=5cm]{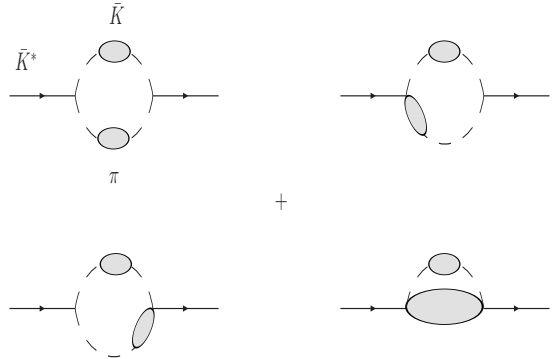}
\hfill
\caption{Self-energy diagrams from the decay of the $\bar{K}^*$ meson in the medium.}
\label{fig:1}
\end{center}
\end{figure}

\begin{figure}[ht]
\begin{center}
\includegraphics[width=0.8\textwidth]{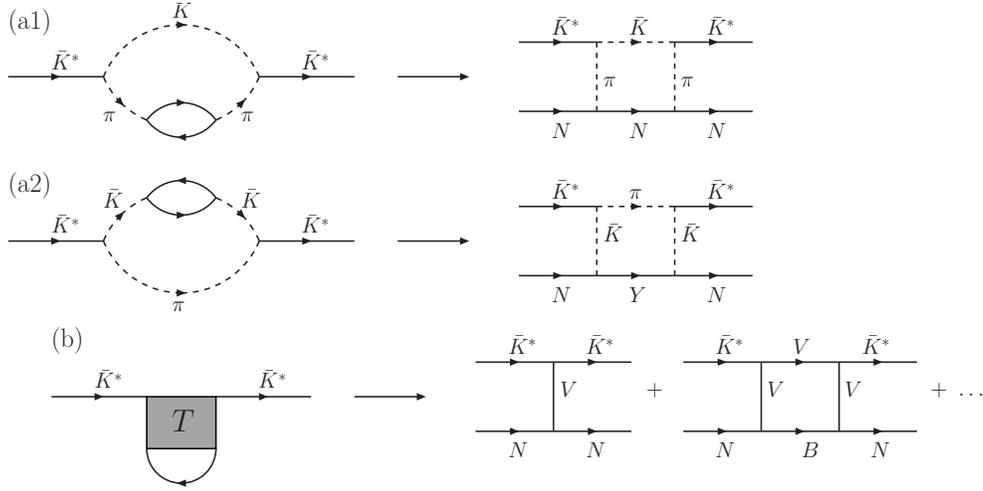}
\caption{Contributions to the $\bar K^*$ self-energy, depicting their different
inelastic sources.}
\label{fig:3}
\end{center}
\end{figure}

The $\bar{K}^*$ self-energy  in matter, on one hand, results from its decay into ${\bar K}\pi$, $\Pi_{\bar{K}^*}^{\rho,{\rm (a)}}$, including both the self-energy of the antikaon \cite{Tolos:2006ny} and the pion \cite{Oset:1989ey,Ramos:1994xy} (see first diagram of Fig.~\ref{fig:1} and some specific contributions in diagrams $(a1)$ and $(a2)$ of Fig.~\ref{fig:3}). Moreover, vertex corrections required by gauge invariance are also incorporated, and they are associated to the last three diagrams in  Fig. \ref{fig:1}. 
Another contribution to the $\bar K^*$ self-energy comes from its interaction
with the nucleons in the Fermi sea, as displayed in diagram (b) of 
Fig.~\ref{fig:3}. This accounts for the direct quasi-elastic process $\bar K^* N \to \bar K^* N$, as well as other absorption channels $\bar K^* N\to \rho Y, \omega Y, \phi Y, \dots$ with $Y=\Lambda,\Sigma$. 
This contribution is determined by
integrating the medium-modified $\bar K^* N$ amplitudes, $T^{\rho,I}_{\bar
K^*N}$, over the  Fermi sea of nucleons and, therefore, it will be sensitive to the resonant structures in these amplitudes.
In particular, the in-medium amplitudes
retain clear traces of two resonances that are generated from the $\bar K^* N$ interaction and related channels
at 1783 MeV and 1830 MeV \cite{angelsvec}, which
can be identified with the experimentally observed $J^P = 1/2^-$ states $\Lambda(1800) $and $\Sigma(1750)$, respectively. Note that the
self-energy $\Pi_{\bar{K}^*}^{\rho,{\rm (b)}}$ has to be determined self-consistently
since it is obtained from the in-medium amplitude
${T}^\rho_{\bar K^*N}$ which contains the $\bar K^*N$ loop function
${G}^\rho_{\bar K^*N}$, and this last quantity itself is a function of the complete self-energy
$\Pi_{\bar K^*}^{\rho}=\Pi_{\bar{K}^*}^{\rho,{\rm (a)}}
+\Pi_{\bar{K}^*}^{\rho,{\rm (b)}}$.

The two contributions to the $\bar K^*$ self-energy, coming from
the decay of
$\bar K \pi$ pairs in the medium [Figs.~\ref{fig:3}(a1) and \ref{fig:3}(a2)] or
from the  $\bar K^* N$ interaction [Fig.~\ref{fig:3}(b)] provide different
sources
of inelastic $\bar K^* N$ scattering, which add incoherently in the $\bar K^*$
width. Note that the $\bar K^* N$ amplitudes mediated by intermediate $\bar K  N$ or $\pi Y$ states are not unitarized, in contrast to what is done for the contributions from intermediate $VB$ states. The problem arises because the exchanged pion may be placed on its mass shell, which forces one to keep track of the proper analytical cuts making the iterative process more complicated. A technical solution can be found by calculating the box diagrams
of Figs.~\ref{fig:3}(a1) and \ref{fig:3}(a2), taking all the cuts into account
properly, and adding the resulting $\bar K^* N \to \bar K^* N$ terms to the $VB
\to V^\prime B^\prime$ potential coming from vector-meson exchange, in a
similar way as done for the study of the vector-vector interaction in
Refs.~\cite{raquel,gengvec}. As we saw in the former sections, the generated resonances barely change their position for spin 3/2 and only by a moderate amount in some cases for spin 1/2. Their widths are somewhat enhanced due to the opening of the newly
allowed $PB$ decay channels \cite{garzon}.

\begin{figure}[ht]
\begin{center}
\includegraphics[width=0.45\textwidth]{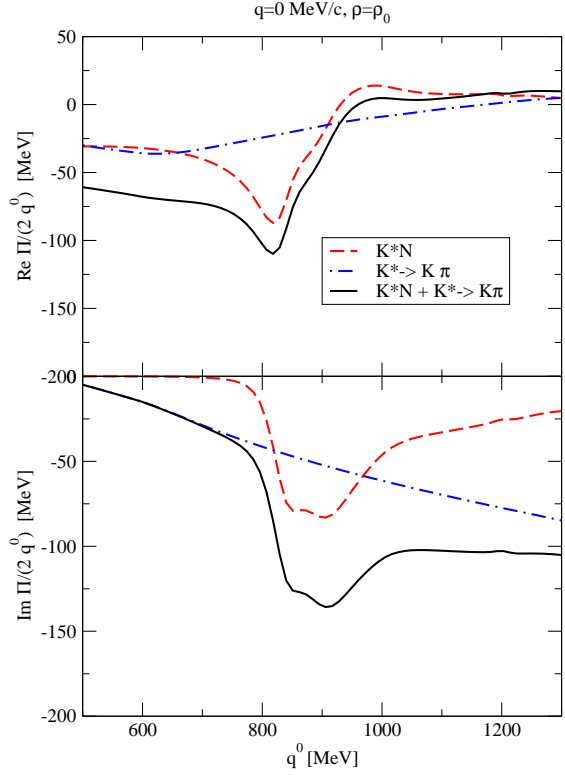}
\hfill
\caption{  $\bar K^*$ self-energy for
 $\vec{q}=0 \, {\rm MeV/c}$ and $\rho_0$. }
\label{fig:auto-spec}
\end{center}
\end{figure}

The full $\bar K^*$ self-energy as a function of the $\bar K^*$ energy for zero
momentum at normal nuclear matter density is shown  in 
Fig.~\ref{fig:auto-spec}. We explicitly indicate the contribution to the
self-energy coming from the self-consistent calculation of the $\bar K^* N$
effective interaction (dashed lines) and the self-energy from the $\bar K^*
\rightarrow \bar K \pi$ decay mechanism (dot-dashed lines), as well as the
combined result from both sources (solid lines).
Around $q^0= 800-900$ MeV we observe an enhancement of the width as well as some structures in the real part of the $\bar K^*$ self-energy. The origin of these structures can be traced back to  the coupling of the $\bar K^*$ to the in-medium $\Lambda(1783) N^{-1}$ and  $\Sigma(1830) N^{-1}$ excitations, which dominate the $\bar K^*$ self-energy in this energy region. However, at lower energies where the $\bar K^* N\to V B$ channels 
are closed, or at large energies beyond the resonance-hole excitations,
the width of the $\bar K^*$ is governed by the $\bar K \pi$ decay mechanism in dense matter.

As we can see, around $q^0=m_{\bar{K}^*}$ the $\bar K^*$ feels a moderately attractive optical potential and acquires a width of $260$ MeV, which is about five times its width in vacuum. A method to measure this large width experimentally was devised in  
\cite{lauraraquel} suggesting to use the transparency ratio defined as
\begin{equation}
T_{A} = \frac{\tilde{T}_{A}}{\tilde{T}_{^{12}C}} \hspace{1cm} ,{\rm with} \ \tilde{T}_{A} = \frac{\sigma_{\gamma A \to K^+ ~K^{*-}~ A'}}{A \,\sigma_{\gamma N \to K^+ ~K^{*-}~N}} \ .
\end{equation}
The quantity $\tilde{T}_A$ is the ratio of the nuclear $K^{*-}$-photoproduction cross section
divided by $A$ times the same quantity on a free nucleon. This describes the loss of flux of $K^{*-}$ mesons in the nucleus and is related to the absorptive part of the $K^{*-}$-nucleus optical potential and, therefore, to the $K^{*-}$ width in the nuclear medium.  In order to remove other nuclear effects not related to the absorption of the $K^{*-}$, it was also suggested to look at  this ratio with respect to that of a medium nucleus like $^{12}$C, $T_A$ (see \cite{luismagas}). Results for $T_A$ as a function of $A$ can be seen in 
\cite{lauraraquel}, indicating a sizable depletion of $\bar K^*$ production in nuclei that we would like to encourage to be measured.

\section{$J/\psi$ suppression}
\label{supp}
$J/\psi$ suppression in nuclei has been a hot topic \cite{Vogt:1999cu}, among other reasons for its possible interpretation as a signature of the formation of quark gluon plasma in heavy ion reactions \cite{Matsui:1986dk}, but many other interpretations have been offered  \cite{Vogt:2001ky,Kopeliovich:1991pu,Sibirtsev:2000aw}. In a recent paper \cite{raquelxiao} a study has been done of different $J/\psi N$ reactions which lead to $J/\psi$ absorption in nuclei. The different reactions considered are 
the transition of  $J/\psi N$ to $VN$ with $V$ being a light vector, $\rho, \omega,\phi$, together with the inelastic channels, 
$J/\psi N \to \bar D \Lambda_c$ and $J/\psi N \to \bar D \Sigma_c$. 
Analogously, the mechanisms where an exchanged $D$ collides with a nucleon giving rise to $\pi \Lambda_c$ or $\pi \Sigma_c$ states are also considered. 

\begin{figure}[ht]
\begin{center}
\includegraphics[width=0.6\textwidth]{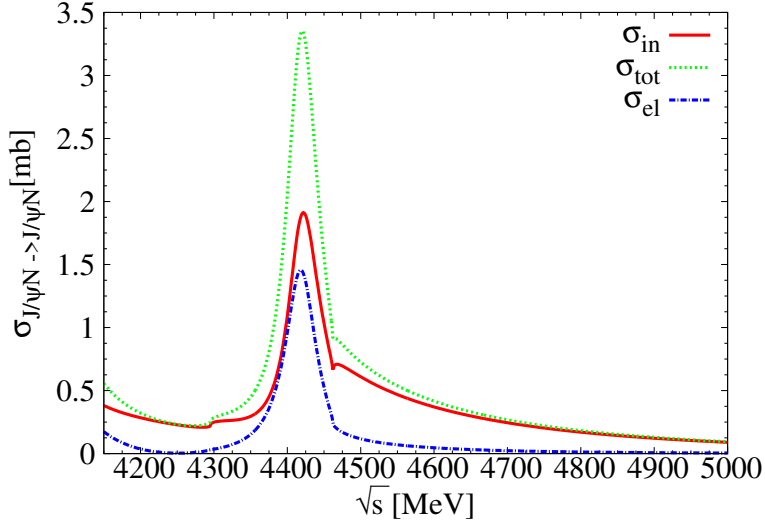}
\caption{The total, elastic and inelastic cross sections obtained from the unitary $J/\psi N \to J/\psi N$ amplitude involving only intermediate vector-baryon states.}\label{crosec}
\end{center}
\end{figure}

The total, elastic and inelastic cross sections, obtained from the unitarized $J/\psi N \to  J/\psi N$ amplitude where only intermediate vector-baryon states are considered, are shown in Fig.~\ref{crosec}. We can clearly see the peak around 4415 MeV produced by the hidden charm resonance, described in Tables \ref{jpsicoupling} and \ref{jpsiwidth}, dynamically generated from the interaction of $J/\psi N$ with its coupled $VB$ channels. 

\begin{figure}[ht]
\begin{center}
\includegraphics[width=0.45\textwidth]{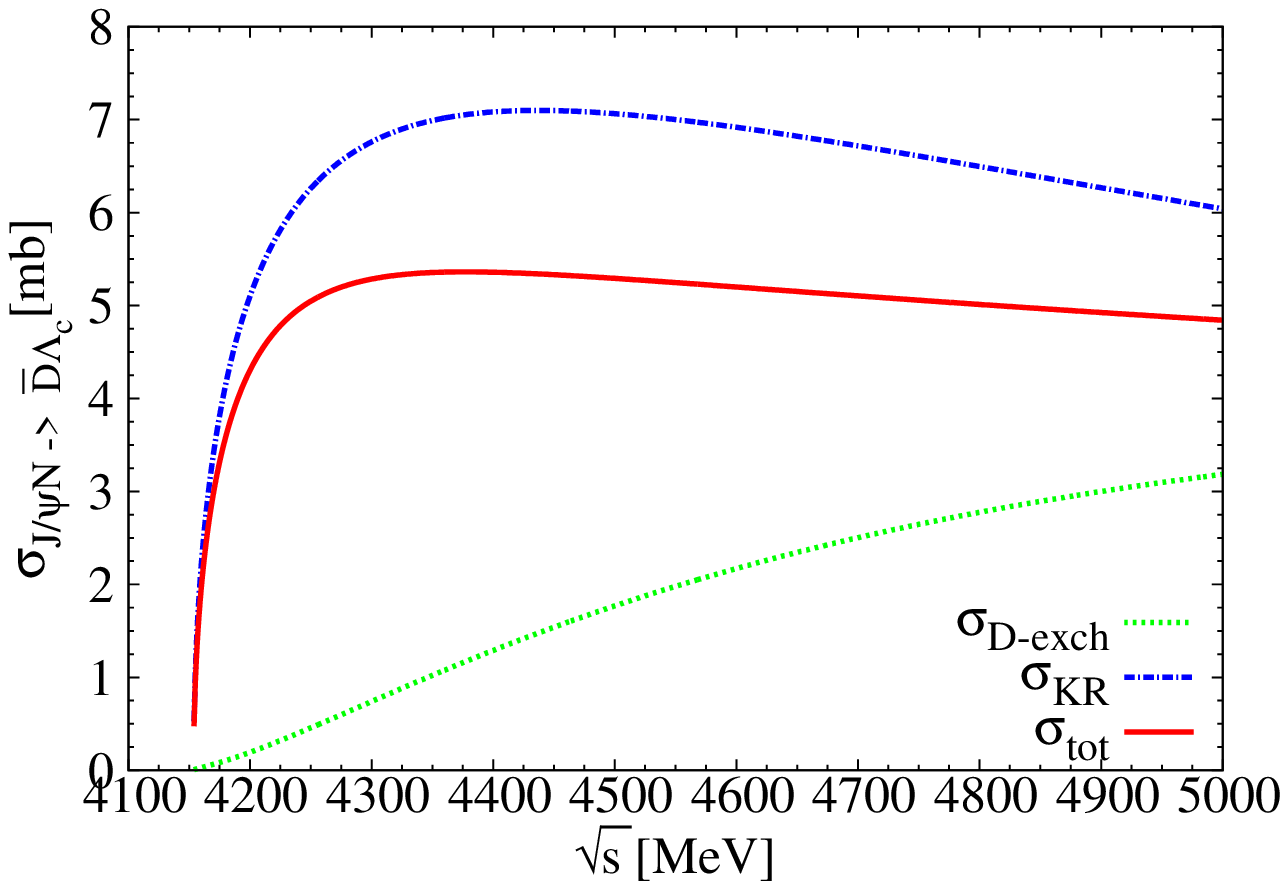}
\includegraphics[width=0.45\textwidth]{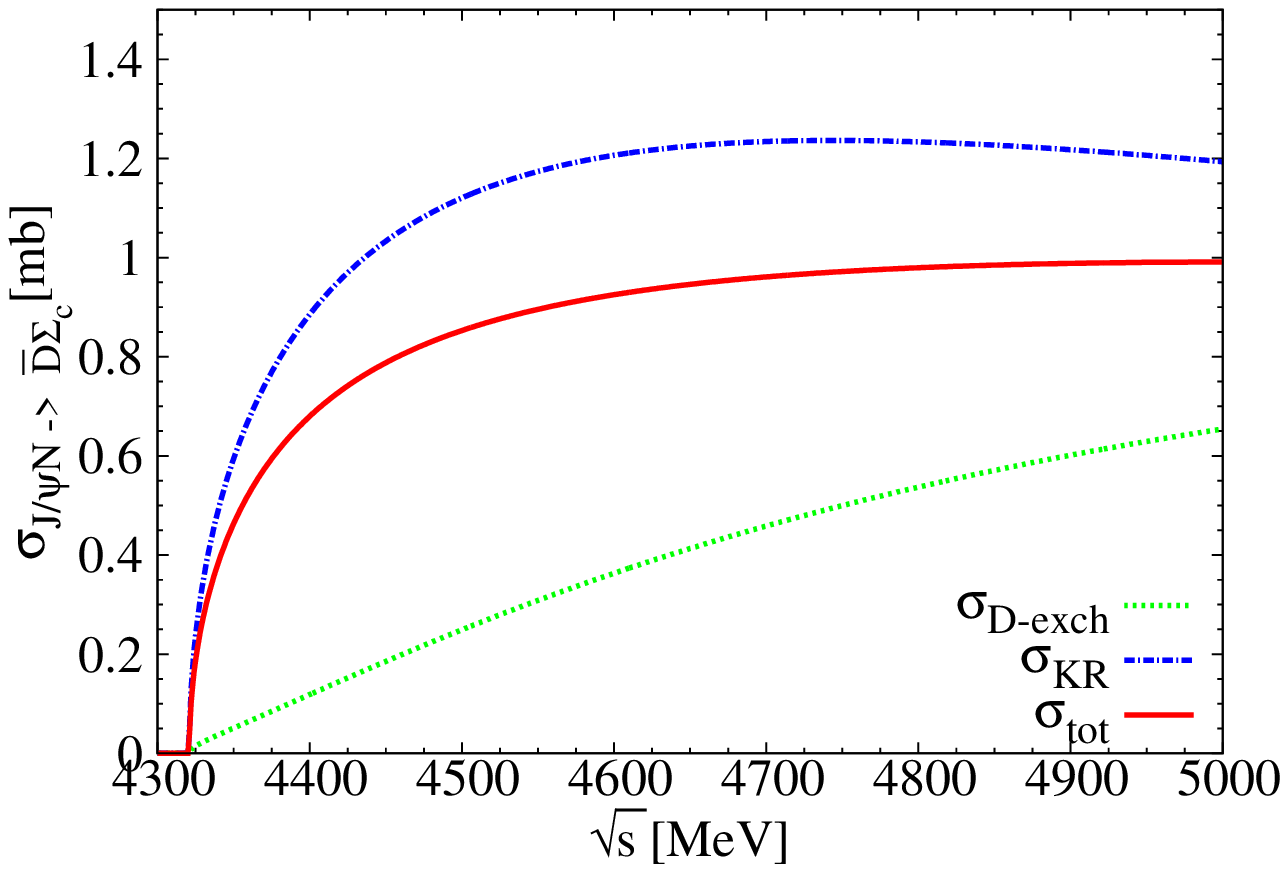}
\caption{The cross section for $J/\psi N \to \bar{D} \Lambda_c$ (up) and $J/\psi N\to \bar{D} \Sigma_c$ (down).}\label{figcro}
\end{center}
\end{figure}

\begin{figure}[ht]
\begin{center}
\includegraphics[width=0.45\textwidth]{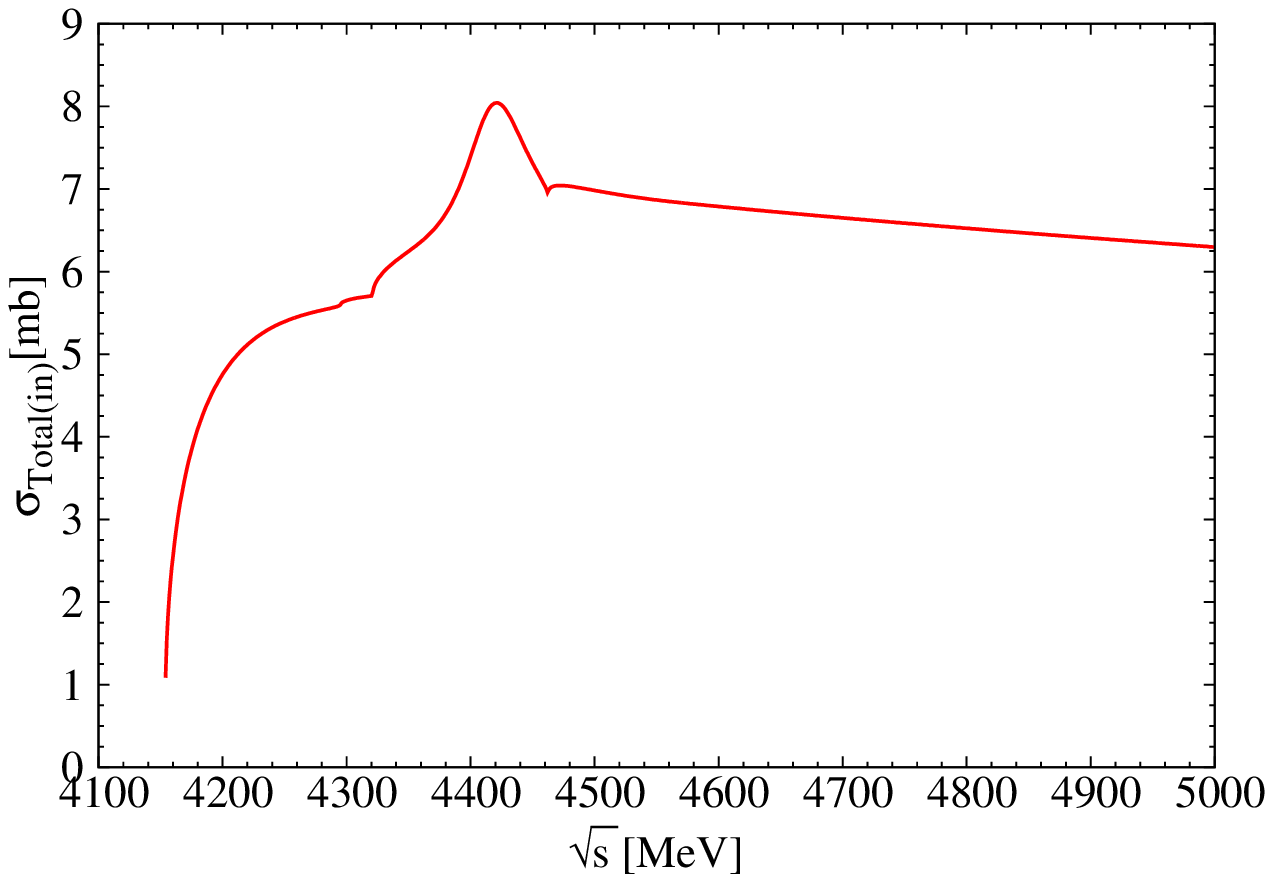}
\includegraphics[width=0.45\textwidth]{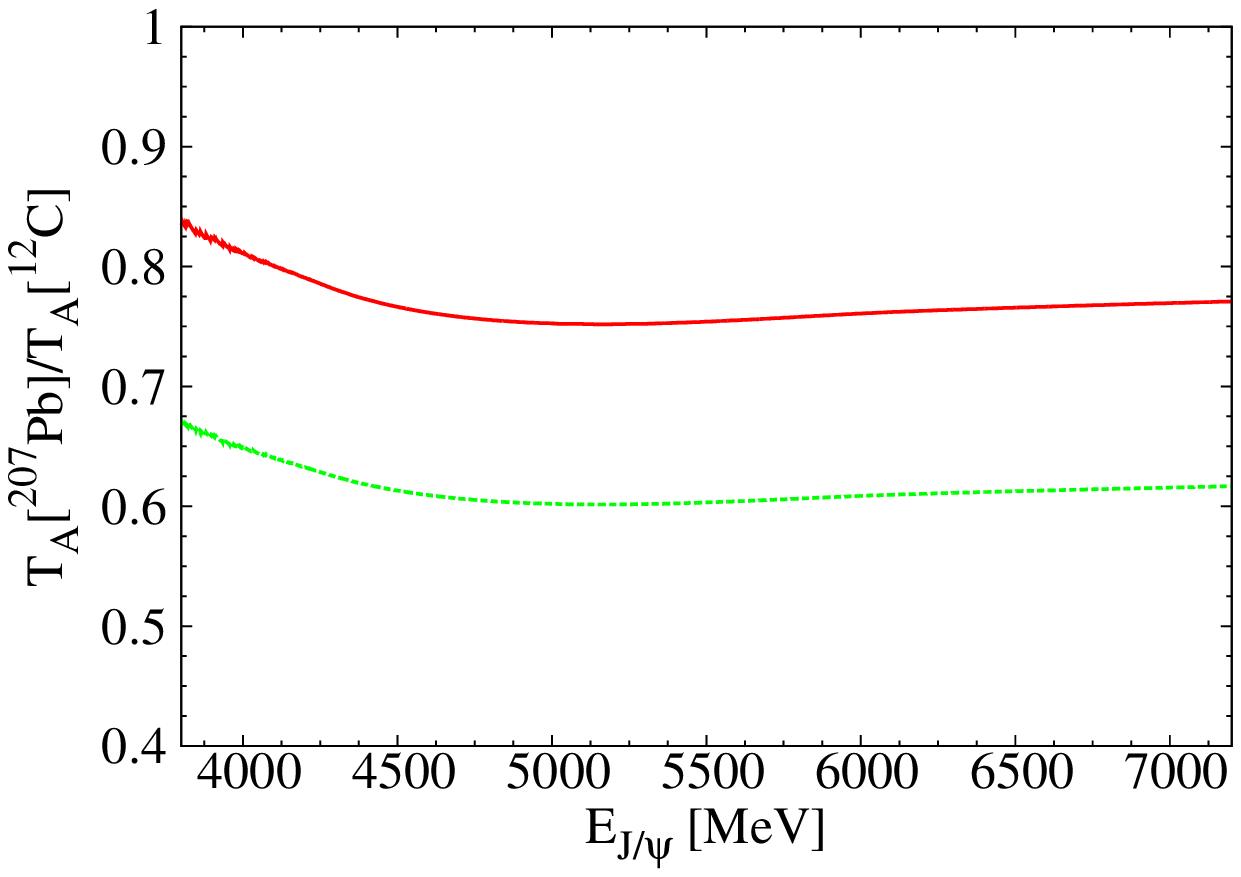}
\caption{Left: The total inelastic cross section of $J/\psi N$. Right: The transparency ratio of $J/\psi$ photoproduction  as a function of the energy in the CM of $J/\psi$ with nucleons of the nucleus. Solid line: represents  the effects due to $J/\psi$ absorption. Dashed line: includes photon shadowing \cite{Bianchi:1995vb}.}\label{sigin}
\end{center}
\end{figure}

The cross sections for the inelastic transitions $J/\psi N \to \bar{D} \Lambda_c$  and $J/\psi N\to \bar{D} \Sigma_c$ are shown in Fig. \ref{figcro}. We can see that the first cross section is sizable and bigger than the one from the $VB$ channels. The cross sections for $J/\psi N  \to \bar D \pi \Lambda_c$ or $\bar D \pi \Sigma_c$ are small in size in the region of interest and are not plotted here. 
The total $J/\psi N$ inelastic cross section, obtained as the sum of all inelastic cross sections from the different sources discussed before, is represented in Fig. \ref{sigin} (left).

With the inelastic cross section obtained, the transparency ratio for electron induced $J/\psi$ production in nuclei at beam energies around 10 GeV has been studied. The results are shown in Fig. \ref{sigin} (right) where the transparency ratio of $^{208}$Pb relative to that of $^{12}$C is displayed as a function of the energy. It is clear that one finds sizable reductions in the rate of $J/\psi$ production in electron induced reactions. 
It should be noted that the calculation of the transparency ratio discussed so far does not consider the shadowing of the photons and assumes they can reach every point without being absorbed. However, for $\gamma$ energies of around 10 GeV, as suggested here, the photon shadowing, or initial photon absorption, cannot be ignored. Taking this into account is easy since one must multiply the ratio $T_A$ by the ratio of $N_{rm eff}$ for a nucleus of mass $A$ relative to $^{12}$C. This ratio for $^{208}$Pb to $^{12}$C at $E_\gamma =$10 GeV is of the order $0.8$, but with uncertainties \cite{Bianchi:1995vb}. This factor is applied to the lower curve of Fig. \ref{sigin} (right) for a proper comparison with experiment. The results for the transparency ratio imply that $30 - 35$ \% of the $J/\psi$ produced in heavy nuclei are absorbed inside the nucleus. This is very much in line with depletions of $J/\psi$ in matter observed in other reactions and offers another perspective in the interpretation of the $J/\psi$ suppression in terms of hadronic reactions, which has also been advocated before \cite{Sibirtsev:2000aw}. Apart from novelties in the details of the calculations and the reaction channels considered, we find that the presence of the resonance that couples to $J/\psi N$ produces a peak in the inelastic $J/\psi N$ cross section and a dip in the transparency ratio. However, this dip is washed away when effects of Fermi motion are taken into account. 

\section{Conclusions}

We have made a survey of recent developments along the interaction of vector mesons with baryons and the properties of some vector mesons in a nuclear medium.  We showed that the interaction is strong enough to produce resonant states which can qualify as quasibound states of a vector meson and a baryon in coupled channels. This adds to the wealth of composite states already established from the interaction of pseudoscalar mesons with baryons. At the same time we reported on studies of the mixing of the pseudoscalar-baryon states with the vector-baryon states which break the spin degeneracy that the original model had.  The mechanisms of vector-baryon interaction extended to the charm sector also produced some hidden charm states which couple to the $J/\psi N$  channel and had some repercussion in the $J/\psi$ suppression in nuclei. We also showed results for the spectacular renormalization of the $K^*$ in nuclei, where the width becomes as large as 250 MeV at normal nuclear matter density and we made suggestions of experiments that could test this large change.

\section*{Acknowledgments}  
This work is partly supported by DGICYT contract numbers
FIS2011-28853-C02-01, FIS2011-24154, the Generalitat Valenciana in the program Prometeo, 2009/090 and
Grant No. 2009SGR-1289 from Generalitat de Catalunya. L.T. acknowledges support from Ramon y Cajal Research Programme, and from FP7-PEOPLE-2011-CIG under contract PCIG09-GA-2011-291679. We acknowledge the support of the European Community-Research Infrastructure
Integrating Activity
Study of Strongly Interacting Matter (acronym HadronPhysics3, Grant Agreement
n. 283286)
under the Seventh Framework Programme of EU.

\end{document}